\documentclass[10pt]{article}
\usepackage[cp1251]{inputenc}
\usepackage[russian]{babel}
\usepackage{graphicx}

\begin{document}

\twocolumn[\noindent{\small\it  ISSN 1063-7729, Astronomy Reports,
Vol. 51, No. 11, 2007, pp. 891--902. \copyright Pleiades
Publishing, Ltd., 2007. \noindent Original Russian Text \copyright
A.T. Bajkova, 2007, published in Astronomicheski$\check{\imath}$
Zhurnal, 2007, Vol. 84, No. 11, pp. 984--996. }

\vskip -4mm

\begin{tabular}{llllllllllllllllllllllllllllllllllllllllllllllll}
 & & & & & & & & & & & & & & & & & & & & & & & & & & & & & & & & & & & & & & & \\
\hline \hline
\end{tabular}

\vskip 1.5cm

\centerline{\LARGE\bf Reconstructing Images from Projections
}\centerline{\LARGE\bf Using the Maximum-Entropy Method.
}\centerline{\LARGE\bf Numerical Simulations of Low-Aspect
Astrotomography }

\bigskip

\centerline{\bf\large  A. T. Bajkova}

\medskip

\centerline{\it Main Astronomical Observatory, Russian Academy of
Sciences, St.Petersburg, 196140 Russia} \centerline{\small
Received February 2, 2007; in final form, April 5, 2007}

\vskip 0.5cm

{\bf Abstract} --- { The reconstruction of images from a small
number of projections using the maximum-entropy method (MEM) with
the Shannon entropy is considered. MEM provides higher-quality
image reconstruction for sources with extended components than the
H$\ddot{o}$gbom CLEAN method, which is also used in low-aspect
astrotomography. The quality of image reconstruction for sources
with mixed structure containing bright, compact features embedded
in a comparatively weak, extended base can be further improved
using a difference-mapping method, which requires a generalization
of MEM for the reconstruction of sign-variable functions. We draw
conclusions based on the results of numerical simulations for a
number of model radio sources with various morphologies. }

\bigskip

PACS numbers : 95.75.Mn, 95.75.Pq, 98.70.Dk, 98.54.Gr

{\bf DOI}: 10.1134/S1063772907110030

\vskip 1cm

]

\centerline{1.~INTRODUCTION }

\medskip

The reconstruction of images from projections (reconstructive
tomography) is of interest in many fields, from medicine to
astronomy [1]. Examples in astronomy include reconstructing the
brightness distributions of astrophysical objects from
lunar-occultation observations [2–4] and Doppler-tomography
analysis of binary star systems [5–8]. The synthesis of images
using radar methods is also of considerable interest [9, 10].

The problem of reconstructing images using nonlinear algorithms
possessing high interpolating and extrapolating properties arises
when the number of projections is small (low-aspect
astrotomography), and there are large unfilled regions (“holes”)
in the data plane. In such cases, the use of linear methods, such
as reverse filtered projections [1], becomes inexpedient due to
the impossibility of reconstructing the missing spectral
information and the poor quality of the resulting images [11].

One class of image-reconstruction methods is based on the
radio-astronomy approach [11–13], using the well known CLEAN
algorithm of H$\ddot{o}$gbom [14] or its modification [15]. In
this case, the observational data (projections) are used to form a
“dirty” image, which is the convolution of the desired
distribution and the synthesized beam. As a result, the task of
reconstructing an image reduces to solving for the inverse
convolution, which is an ill-posed problem.

Agafonov and Podvoiskaya [11, 12] propose solving this problem as
is traditionally done in aperture synthesis. In particular, they
use an algorithm with two CLEAN procedures in order to obtain two
solutions, one of which better represents the compact and the
other the extended features in the source structure [12]. The
first solution is obtained by applying a standard H$\ddot{o}$gbom
CLEAN, and the second using the modified CLEAN algorithm proposed
in [15] for reconstructing images of extended sources. However,
methods that are able to reconstruct all of the structural
components of a source equally accurately, thus yielding a more
objective representation of its structure, are of greater
interest. One such alternative image-reconstruction method is the
maximum-entropy method (MEM), using the Shannon entropy [16, 17],
which is also well known in low-aspect astrotomography [5–7].

The absence of any {\it a priori} restrictions on the source
structure (except for the assumption that it is finite in extent)
makes MEM a more fundamental method than the CLEAN method
initially proposed for the reconstruction of point-like sources.
Another advantage of MEM over both the standard and modified [15]
CLEAN methods is its independence of a number of specified
parameters, whose choice appreciably influences the quality of the
reconstruction.

The aim of the current work is to investigate the possibilities of
MEM in the reconstruction of images with various types of
morphologies when the number of available projections of the data
is very limited. We also consider ways to improve MEM to make it
more suitable for the reconstruction of images consisting of
bright, compact features embedded in a weaker, extended base. To
solve this latter problem, we propose a difference maximum-entropy
method, whose realization requires a generalization of the
standard MEM algorithm to the case of reconstructing sign-variable
functions.

In the following sections of the paper, we present a mathematical
formulation of low-aspect astrotomography, describe the process of
reconstruction from data projections, describe the maximum-entropy
method (both its generalized form and the difference method), and
carry out a comparative analysis of the results of numerical
simulations of image reconstruction.

\vskip 1cm

\centerline{2.~FORMULATION OF THE PROBLEM}

\medskip

The problem we consider here can be formulated as follows. Let us
consider a two-dimensional source of radiation having a finite
spatial extent on the sky in the region $(X, Y)$. Such an object
is described mathematically by a two-dimensional, finite,
non-negative function of the variables $X$ and $Y$ , $O(X, Y)$
(Fig. 1a). The projection of the object onto a line $p$ making an
angle $\phi$ with the $Y$ axis, as is shown in Fig. 1a, is the
integral of the intensity distribution in the object over the
coordinate perpendicular to the direction of $p$. The projection
is a function of the variable $p$,which we denote $T(p)$. Let
there be $N$ projections of the object at different angles
$\phi_i$, so that there are $N$ integrals $T_i(p) (i=1,…,N)$ for
the two-dimensional object along directions comprising angles
$\phi_i+90^{\circ}$ with the $Y$ axis. Our task is to reconstruct
the image $O(X,Y)$ from the projections $T_i(p)$.

\vskip 1cm

\centerline{3. RECONSTRUCTION FROM PROJECTIONS}

\medskip

At the basis of the algorithm we used to reconstruct images from
projections lies the fundamental relation between the Radon
transform and the Fourier transform, formulated as a projection
theorem [1]: the Fourier transform of a projection $T(p)$ at an
angle $\phi$ to the $Y$ axis (Fig. 1a) is the one-dimensional
central cross section of the two-dimensional Fourier transform of
the function $O(X,Y)$ at the same angle $\phi$ to the $V$ axis in
the spatial-frequency domain $(U, V)$ (the $UV$ plane). Figure 1b
shows the $(U, V)$ coordinates of the Fourier transform of the
projection $T(p)$. An example with six projections, which we used
for our simulations, is illustrated in Fig. 2a, and the
corresponding coverage of the UV plane is shown in Fig.2b. If the
number of projections is infinite, the reconstruction of the image
is obtained via an inverse Radon transform [1].

Note that the fundamental possibility of reconstructing an image
from a small number of projections, or equivalently with
incomplete coverage of the $UV$ spatial-frequency domain, is based
on the property that the Fourier transform of the finite function
describing the spatially finite object is analytic [18]. Analytic
functions can be extended throughout an infinite space in which
they are defined using the known values in a finite interval or at
a finite set of points.

The main requirement for a reconstruction algorithm is
non-linearity. Only due to this non-linearity is it possible to
realize the “analytic continuation” of a spectrum; i.e., to fill
empty regions (“holes”) in the UV plane [17]. It is fundamentally
impossible to obtain the missing spectral harmonics using linear
procedures. In many cases, the requirement that the solution be
non-negative turns linear procedures into appreciably non-linear
ones (for example, in the least-squares method). In the standard
MEM approach, non-negativity is an intrinsic internal property of
the solution, indicating the appreciable non-linearity of the
method.

\vskip 1cm

\centerline{4. THE MAXIMUM-ENTROPY METHOD}

\medskip

MEM is one of a large class of non-linear informational methods
[19] whose essence lies in the optimization of a functional
specified by some informational criterion for the quality of the
solution (the maximum information, entropy, $\alpha$- divergence,
etc.), subject to the fulfillment of various linear or nonlinear
constraints that flow from the data [17]. In our case, maximizing
the Shannon entropy consists in finding the maximum of the
functional

\begin{equation}
E = \int x(t) \ln(1/x(t)) dt = -\int x(t) \ln(x(t)) dt,
\end{equation}
where $x(t)$ is the desired distribution and, in accordance with
the projection theorem, measurements of the spatial Fourier
spectrum of the source serve as the data to be used, leading to
linear constraints in the form of equalities.

Since the numerical simulations suppose working with digital data,
we present a discrete formulation of the optimization. Let a map
of an object with a finite carrier be discretized in accordance
with the Shannon–Kotel’nikov theorem and have a size of $N\times
N$ pixels. We denote the discrete measurements of the desired
distribution

$$
x_{kl},~~~~ k,l=1,...,N-1.
$$
We denote the known measurements of the two - dimensional Fourier
spectrum of the object, which represent the data, in accordance
with the projection theorem, as follows, separating the real,
$A_m$, and imaginary, $B_m$, parts:

$$
X_m=A_m+jB_m,~~~~m=1,...,M,
$$
where $M$ is the number of known measurements and $m$ is the
number of the current measurement with coordinates $(u_m,v_m)$  in
the $UV$ plane, not necessarily located at nodes of the coordinate
grid. This last circumstance means that there is no problem with
pixelization of the data in the frequency domain, which represents
a certain technical advantage of this method over other methods
[1] and appreciably enhances the accuracy of the reconstruction.

The practical MEM algorithm we applied, taking into account the
errors in the data [17, 20], supposes the solution of the
conditional-optimization problem

\begin{equation}
\min \sum_k\sum_l
x_{kl}\ln(x_{kl})+\rho\sum_m\frac{(\eta_m^{re})^2+(\eta_m^{im})^2}{\sigma_m^2},
\end{equation}

\begin{equation}
\sum_k\sum_l x_{kl}a_{kl}^m-\eta_m^{re}=A_m,
\end{equation}

\begin{equation}
\sum_k\sum_l x_{kl}b_{kl}^m-\eta_m^{im}=B_m,
\end{equation}

\begin{equation}
x_{kl}\ge 0,
\end{equation}
where $a_{kl}^m$ and $b_{kl}^m$ are constant coefficients (cosines
and sines) that flow from the Fourier transform, $A_m$ and $B_m$
are the real and imaginary parts of the spectral data for the
object, $\eta_m^{re}$ and $\eta_m^{im}$ are the real and imaginary
parts of the instrumental additive noise, which has a normal
distribution with zero mean and known dispersions $\sigma_m$, and
$\rho$ is a positive weight.

As we can see from (2), the optimized functional has two parts: a
Shannon-entropy functional and a functional that is an estimate of
the difference between the reconstructed spectrum and the measured
data according to a $\chi^2$ criterion. This latter functional can
be considered an additional regulating, or stabilizing, term
acting to provide a further regularization of the solution above
that possible with the entropy functional alone [20]. The
influence of this additional term on the resolution of the
reconstruction algorithm must be borne in mind [7].

Equations (3)–(4) represent linear constraints on the unknown
images $x_{kl}$ and noise terms $\eta_m^{re}$ and $\eta_m^{im}$.
The non-negativity constraint on the image (5) can be omitted in
this case due to the nature of the entropy solution, which is
purely positive.

Note that, due to the fact that there is always a zero point in
the $UV$ plane (the total flux of the source $F_0$), this
automatically brings about the normalization of the solution
required for MEM:

$$
\sum_k\sum_l x_{kl} = F_0.
$$

The numerical algorithm for the reconstruction (2)–(5) treating
this as a non-linear optimization problem based on the
Lagrange-multiplier method, is considered in detail in [19]. Here,
we present only the solution:
\begin{equation}
x_{kl}=\exp(-\sum_m(\alpha_m a_{kl}^m+ \beta_m b_{kl}^m)-1),
\end{equation}

\begin{equation}
\eta_m^{re} = \frac{\sigma_m^2 \alpha_m}{\rho}, ~~~~\eta_m^{im} =
\frac{\sigma_m^2 \beta_m}{\rho},
\end{equation}
expressed in terms of the Lagrange multipliers (dual
variables)$\alpha_m$ and $\beta_m$, through which the constraints
(3) and (4), respectively, enter the Lagrange functional.

As we can see from (6), the standard MEM image is manifestly
positive. It is not difficult to show that the MEM Hesse matrixes
everywhere positive definite, so that the entropy functional is
convex and the solution is global.

Various gradient methods can be used to search for the extrema of
the corresponding dual functional. We used a coordinate-descent
method, since it is the most reliable means to search for the
global solution.

\vskip 1cm

\centerline{5. GENERALIZED MAXIMUM-ENTROPY METHOD}

\medskip

The realization of the difference-mapping principle described in
Section 6 requires a generalized MEM algorithm for the
reconstruction of arbitrary real functions that take on both
positive and negative values. We adopt the following definition of
the entropy of such an arbitrary real function.

{\bf Definition 1.} The entropy E of an arbitrary real function
$x(t)$ that can, in general, take on both positive and negative
values is defined as the entropy of the modulus of this function,
$|x(t)|$ [20]:
\begin{equation}
\label{5} E = -\int |x(t)| \ln(|x(t)|) dt.
\end{equation}

We denote the positive part of the function $x(t)$ as $x^p(t)$,
and its negative part $-x^n(t)$, with both $x^p(t)$ and $x^n(t)$
being non-negative: $x^p(t), x^n(t) \ge 0$. The function x(t) can
then be represented in the form
\begin{equation}
\label{6} x(t) = x^p(t) - x^n(t),
\end{equation}
with $x(t)$ being determined by only one of these terms at each
value $t$ — either $x^p(t)$ or $x^n(t)$; in this case, we can
define the modulus of the function
\begin{equation}
\begin{array}{ccc}
x(t) = ~~x^p(t)~~~ {\hbox {\rm если}} ~~~ x(t) \ge 0,\\
 & & \\
x(t) = -x^n(t)~~ {\hbox {\rm если}} ~~~ x(t) < 0.
\end{array}
\end{equation}
Obviously, the relation
\begin{equation}
x^p(t) \cdot x^n(t) = 0
\end{equation}
is equivalent to the conditions (10).When the conditions (10) are
fulfilled, expression (8) takes the standard form
$$
E = -\int x^p(t) \ln(x^p(t)) + x^n(t) \ln(x^n(t)) dt.
$$

Let us also define the associated entropy [20].

{\bf Definition 2.} The associated entropy of the function $x(t)$
is given by the expression
\begin{equation}
E(\alpha)= - \int x(t) \ln(\alpha x(t)) dt,
\end{equation}
where $\alpha$ is a real, positive parameter. The associated
entropy (12) differs from the usual entropy (1) in the presence of
a in the logarithm.

Let us now define the generalized entropy of a real, sign-variable
function.

{\bf Definition 3.} The generalized entropy of a real,
sign-variable function (9) is its associated entropy, calculated
as follows:
\begin{equation}
E(\alpha)= -\int (x^p(t) \ln(\alpha x^p(t)) + x^n(t) \ln(\alpha
x^n(t)))~dt.
\end{equation}
In this case, the parameter $\alpha$  plays a leading role, since
its value determines the accuracy with which the positive and
negative parts of the solution $x(t)$ can be separated, and thus
the quality of the reconstructed image.

Let us consider this further. Omitting all intermediate steps, we
write the solution for the corresponding discrete optimization
problem (13) in terms of the dual variables:

\begin{equation}
\begin{array}{ccc}
x_{kl}^p & = & \exp(-\sum_m(\alpha_m a_{kl}^m + \beta_m b_{kl}^m)-1-\ln \alpha),\\
 & & \\
x_{kl}^n & = & \exp(\sum_m(\alpha_m a_{kl}^m + \beta_m b_{kl}^m)-1-\ln \alpha),\\
\end{array}
\end{equation}
whence it follows that the series $x_{kl}^p$ and $x_{kl}^n$ are
related by an expression that depends only on $\alpha$:
\begin{equation}
\label{15} x_{kl}^p \cdot x_{kl}^n = \exp(-2-2\ln \alpha) =
K(\alpha).
\end{equation}

As we can see from (15), $\alpha$ indeed plays the role of an
agent separating the positive and negative parts of the solution
(14). We can achieve any specified accuracy with which the
conditions (10) and (11) are to be satisfied by varying $\alpha$,
since $K(\alpha) \to 0$ as $\alpha \to \infty$. However, we note
that the value of $\alpha$ is bounded from above by purely
computational effects. In our simulations, $\alpha$ was specified
to be 1000. When the noise is taken into account as is done in
(2)–(4), the solution (7) also appears.

\vskip 1cm

\centerline{6. DIFFERENCE-MAPPING METHOD}

\medskip

The difference-mapping method is based on the fundamental
linearity of the Fourier transform. Bright components in the
source that are reconstructed in the first stage are subtracted
from the input spectrum, the remaining reconstruction is carried
out for the residual spectral data, and the results of the two
reconstructions are finally summed.

In the following section, we show that, when MEM $-$ a method with
clearly expressed non-linear properties $-$ is used for the
reconstruction, the difference-mapping method can lead to an
appreciable improvement in the quality of the resulting image,
especially when the source has both compact features and a fairly
weak extended base. This type of source structure is most
problematic from the point of view of the accuracy with which all
the structural components are reconstructed. The reconstruction of
uniform structures $-$ either purely compact or purely extended
$-$ is much simpler (see also Section 7).

What is the reason for this improvement in the quality of the
reconstruction? After subtracting from the input spectral data
bright components reconstructed by MEM in the first stage, we
obtain a residual spectrum in which the weak, extended component
comprises a high proportional compared to the compact component.
(We will call the result of this subtraction the first-order
residual spectrum.) In this way, we have artificially lowered the
dynamic range and increased the structural uniformity of the map
corresponding to the residual spectrum, thereby facilitating the
accurate reconstruction of the image in the second stage of the
procedure.

If we subtract from the residual spectrum bright components in the
image reconstructed in the second stage, we will obtain a
second-order residual spectrum, which will correspond to a map
with still lower dynamic range and higher structural uniformity.
Continuing in this way, we can obtain residual functions of the
spectrum of higher and higher order. To obtain the desired map of
the source, we must sum the map reconstructed in the final stage
with all the components subtracted from the input spectrum in
previous stages.

Formally, the two-stage difference-mapping algorithm can be
represented
$$
F_{sp}^1 = F_{sp}^0 - F_{sp}^{br1},
$$
where $F_{sp}^0$ are the input data for the spectrum of the
desired brightness distribution over the source $x(t)$,
$F_{sp}^{br1}$ is the spectrum corresponding to the bright
components reconstructed in the first stage $x(t)_{br1}$, and
$F_{sp}^1$ is the first-order residual spectrum corresponding to
the map of the source reconstructed in the second stage $x(t)_2$.
The resulting map will have the form
$$
x(t) = x(t)_{br1} + x(t)_2.
$$

Let us now note an important property of the difference mapping.
The residual spectral data obtained after subtracting bright
components reconstructed in previous stages of the algorithm can
correspond to an image with negative values. This can occur if
early bright components are reconstructed with overestimated
amplitudes, as is quite probable for any non-linear method. To
avoid undesirable nonlinear distortions of the maps, we must carry
out the reconstruction using the MEM algorithm generalized for
functions that can take both positive and negative values,
described in Section 5.

\vskip 1cm

\centerline{7. RESULTS OF THE NUMERICAL SIMULATIONS}

\medskip

We present here the results of our numerical simulations of
low-aspect astrotomography for six model sources with various
morphologies, denoted {\it1} to {\it6} (Fig. 3). Source {\it1} is
a set of three resolved, compact Gaussian features; source {\it2}
a set of three unresolved, extended Gaussian features; source
{\it3} a combination of extended and compact Gaussian features;
and sources {\it4, 5}, and {\it6} close double sources with a
weak, extended background. Sources {\it4, 5,} and {\it6} have
different intensities for the extended base emission (in {\it4},it
comprises 30\% of the maximum amplitude of the components, and in
{\it5} and {\it6}, only 10\%) and different ratios for the
amplitudes of the compact features (in {\it4} and {\it5}, the
amplitude ratio is 1.0 : 0.8, and in {\it6}, 1.0 : 0.3).

We obtained input data for each numerical image reconstruction
simulation using six projections at angles $\phi_1=0^o$,
$\phi_2=90^o$, $\phi_3=45^o$, $\phi_4=135^o$, $\phi_5=63.5^o$,
$\phi_6=116.5^o$. The arrangement of these projections in the
image plane is shown in Fig. 2a, where the numbers denote the
projection numbers. The synthesized coverage of the $UV$ spatial
frequency plane derived using the projection theorem is shown in
Fig. 2b. A modest amount of instrumental noise having a normal
distribution with a zero mean was added to the artificial data,
such that the signal-to-noise ratio for all examples was about
ten, as is typical for astronomical observations of this type.

The aim of the simulations was to investigation capabilities of
the developed algorithms in reconstructing images of sources with
various structures. This is most interesting from the point of
view of the preferred use of some specific reconstruction
algorithm in some specific situation.

All the model sources shown in Fig. 3 are finite, and their
spectra are analytic, indicating the fundamental possibility of
reconstructing their images using an incomplete set of spectral
data. A small (insignificant) disruption of the analyticity
condition arose due to the addition of a modest amount of noise to
the artificial data. However, as was established in the course of
the simulations, small variations in the data lead to small
variations in the solutions, indicating the stability of the
developed algorithms against noise.

The images were reconstructed from the projections using the
standard MEM algorithm, the generalized difference MEM procedure,
and the H$\ddot{o}$gbom CLEAN algorithm, to enable comparison of
the proposed algorithms with an algorithm traditionally used in
radio astronomical data analysis [11-13].

The results of reconstructing sources {\it 1–6} are presented in
Figs. 4–9; panels (a) show the dirty images, corresponding to the
real coverage of the $UV$ plane (Fig. 2b), calculated as the
inverse Fourier transform of the spectral-data measurements;
panels (b) the images reconstructed using the standard CLEAN
method with loop gain $\gamma$ = 0.1; panels (c) the images
reconstructed using the standard MEM algorithm; and panels (d) the
images reconstructed using the difference-mapping method with the
generalized MEM. In all the images, the minimum contour and the
steps between the contours are 2.5\% of the peak value.

Our analysis of these images shows that the quality of the
reconstruction depends substantially on the structure of the
object. Let us consider the MEM maps (maps (c) in Figs. 4–9). The
highest-quality reconstruction was obtained for uniform sources,
consisting purely of either compact or extended features (sources
{\it1} and {\it2}). In these cases, we obtain a nearly exact
reconstruction (the errors in the reconstruction
are no more than 3\% and no more than 1
respectively). The reconstruction is worse for sources comprised
of a set of extended and compact features (sources {\it3-6}; the
maximum errors are about 10\% for source {\it3} and, on average,
20\% for sources {\it4-6}). The higher the dynamic range of the
map, the lower the quality of the reconstruction, especially for
the extended part of the structure. We were not able to accurately
reconstruct the extended structures in sources {\it4-6}, and we
can see that their outlines repeat the fan-like pattern of the
$UV$ coverage; this demonstrates that it was not possible to
adequately interpolate the spectral data between the sections
shown in Fig. 2b. Moreover, the quality of the reconstruction
becomes worse the weaker the extended base in the source
structure. In this case, the difference mapping method comes to
the rescue. As we can see from maps (d) in Figs. 4-9, the largest
improvement from the difference-mapping method is obtained for
compact structures embedded in a weak, extended base (sources
{\it4–6}). This method was able to reconstruct both the compact
and extended features with high accuracy. The quantitative
characteristics of the reconstructed maps show that applying the
difference method makes it possible to increase the
signal-to-noise ratio and decrease the maximum discrepancy between
the true and reconstructed images by, on average, a factor of
three; the effect of applying the difference method becomes
greater the weaker the extended base in the source structure.

Let us now turn to the CLEAN maps. The images in panels (b) in
Figs. 4-9 show that the best reconstruction of the source shape
occurs for the compact structures (source {\it 1}). However, even
in this case, CLEAN was not able to reconstruct the amplitude
ratios as accurately as MEM: the least compact source is most
stretched in amplitude, and is narrowed in space (the maximum
error in the reconstruction was 30\%). In the case of a purely
extended structure (source {\it 2}), the solution corresponds to a
patchy image, although the general shape of the object was
reproduced reasonably well (the maximum error was about 20\%). The
worst reconstruction was obtained for source {\it 3}, with a mixed
structure. The reconstructed amplitude ratios for the point-like
components were distorted (the maximum error was about 45\%). The
extended part of the structure was also appreciably distorted, and
the area of the extended base was substantially decreased, leading
to large distortions in the general shape of the object. We
observe the same behavior for sources {\it 4–6}. Here, compact
features were reproduced better by the standard MEM (on average,
by 14\%), but there were large spatial distortions in the
reconstruction of the extended base, and we observe a substantial
narrowing of this base, which becomes more prominent the weaker
its intensity.

Thus, these results of our numerical simulations show that the MEM
algorithm is preferable to the traditional CLEAN algorithm,
especially from the point of view of accurately reconstructing
extended source structures. In the case of purely compact sources,
the use of CLEAN is quite justified, especially given the quick
operation of this algorithm. Application of the difference-mapping
method is expedient for sources containing both compact and
comparatively weak extended features, when accurate reconstruction
of all components of the structure is required.

Obviously, the comparative analysis of the various methods we have
presented here is far from complete; in particular, we have not
considered the modified CLEAN algorithm of [15], designed for the
reconstruction of extended sources. However, the comparison of the
standard and modified CLEAN algorithms presented in [12] shows
that the standard CLEAN is best able to reconstruct compact
sources, and the modified CLEAN extended sources. Therefore, in
the case of complex, mixed structures, it was proposed to use a
two-CLEAN method to obtain the entire range of possible solutions.
We have proposed such a MEM algorithm based on a difference
approach that enables the reconstruction of both extended and
compact structures in an object with equal accuracy, demonstrating
its advantages over both the standard and the modified CLEAN
algorithms.

\vskip 1cm

\centerline{8. CONCLUSION}

\medskip

Due to the validity of the projection theorem, the application of
MEM provides a completely natural and simple solution for the
reconstruction of images from a small number of projections.

The results of our comparison of the developed reconstruction
methods based on MEM with the standard CLEAN method demonstrate
the advantages of the former, since they provide a higher quality
reconstruction of extended features. This is due to the fact that
the MEM solutions are smooth, while the standard CLEAN algorithm
operates in a space of $\delta$-functions, leading to
discontinuous solutions. The application of a standard CLEAN
algorithm is justified for compact structures, since it provides a
sufficiently high quality reconstruction and rapid operation in
this case.

The most difficult task is the reconstruction of images with mixed
structures containing bright, compact and weaker extended
features. For this case, we have proposed and studied a difference
mapping method based on the MEM algorithm generalized for the
reconstruction of sign-variable functions. Our numerical
simulations show that this difference-mapping method is able to
appreciably increase the accuracy of the reconstruction of both
extended and compact features, compared to the traditional MEM
algorithm. A comparison of our simulation results with the results
of [12] indicate that the MEM difference-mapping method is
preferable to both the standard CLEAN algorithm and the modified
CLEAN algorithm of [12], designed for the reconstruction of images
of extended sources.

Moreover, the generalized MEM algorithm for the reconstruction of
sign - variable (and also complex) functions [20] makes it
possible to substantially expand the range of practical
application of this algorithm for the reconstruction of various
physical signals.

We note also that there is no problem with the pixelization of
data in the frequency domain in the case of MEM, which is a
technical advantage that appreciably influences the accuracy of
the reconstructions that can be obtained [1].

\vskip 1cm

\centerline{9. ACKNOWLEDGMENTS}

\medskip

The author thanks the referee for valuable comments that led to
improvement of this article.

\vskip 1cm

\centerline{REFERENCES }

\medskip

\noindent 1.~G. Hermen, {\it Image Reconstruction from
Projections} (Academic, New York, 1980; Mir, Moscow, 1983).

\vskip 2mm

\noindent 2.~S. Hoener, Astrophys. J. {\bf 140}, 65 (1964).

\vskip 2mm

\noindent 3.~F. P. Maloney and S. T. Gottesman, Astrophys. J. {\bf
234}, 485 (1979).

\vskip 2mm

\noindent 4.~M. I. Agafonov, V. P. Ivanov, and O. A. Podvoiskaya,
Astron. Zh. {\bf 67}, 549 (1990) [Sov. Astron. {\bf 34}, 275
(1990)].

\vskip 2mm

\noindent 5.~T. R. Marsh and K. Horne, Mon. Not. R. Astron. Soc.
{\bf 235}, 269 (1988).

\vskip 2mm

\noindent 6.~T. R. Marsh and K. Horne, Astrophys. J., {\bf 364},
637 (1990).

\vskip 2mm

\noindent 7.~T. R.Marsh, astro-ph/0011020 v1 (2000).

\vskip 2mm

\noindent 8.~M. I. Agafonov, O. I. Sharova, and M. T. Richards,
Preprint No 505, NIRFI (Radiophysical Research Institute, Nizhnii
Novgorod State University, Nizhnii Novgorod, 2006).

\vskip 2mm

\noindent 9.~A. F. Kononov, Zarub. Radioelektron. No 1, 35 (1991).

\vskip 2mm

\noindent 10.~V. I. Koshelev, S. E. Shipilov, and V. P. Yakubov,
Radiotekh. Elektron. {\bf 44}, 301 (1999) [J. Commun. Technol.
Electron. {\bf 44}, 281 (1999)].

\vskip 2mm

\noindent 11.~M. I. Agafonov and O. A. Podvoiskaya, Izv. Vyssh.
Uchebn. Zaved., Radiofiz. {\bf 32}, 742 (1989).

\vskip 2mm

\noindent 12.~M. I. Agafonov and O. A. Podvoiskaya, Izv. Vyssh.
Uchebn. Zaved., Radiofiz. {\bf 33}, 1185 (1990).

\vskip 2mm

\noindent 13.~M. I. Agafonov and O. I. Sharova, Izv. Vyssh.
Uchebn. Zaved., Radiofiz. {\bf 48}, 367 (2005).

\vskip 2mm

\noindent 14.~J. A. H$\ddot{o}$gbom, Astron. Astrophys., Suppl.
Ser.{\bf 15}, 417 (1974).

\vskip 2mm

\noindent 15.~D. G. Steer, P. E. Dewdney, and M. R. Ito, Astron.
Astrophys. {\bf 137}, 159 (1984).

\vskip 2mm

\noindent 16.~B. R. Frieden, J. Opt. Soc. Am. {\bf 62}, 511
(1972).

\vskip 2mm

\noindent 17.~R. Narayan and R. Nityananda, Ann. Rev. Astron.
Astrophys. {\bf 24}, 127 (1986).

\vskip 2mm

\noindent 18.~Ya. I. Khurgin and V. P.Yakovlev, {\it Finite
Functions in Physics and Engineering} (Nauka, Moscow, 1971) [in
Russian].

\vskip 2mm

\noindent 19.~A. T. Bajkova, Soobshch. Inst. Prikl. Astron., No 58
(IPA RAN, St. Petersburg, 1993).

\vskip 2mm

\noindent 20.~B. R. Frieden and A. T. Bajkova, Appl. Opt. {\bf
33}, 219 (1994).

\bigskip

{\noindent \it Translated by D. Gabuzda }

\onecolumn

\begin{figure}[t]
{\begin{center}
 \includegraphics[width= 100mm]{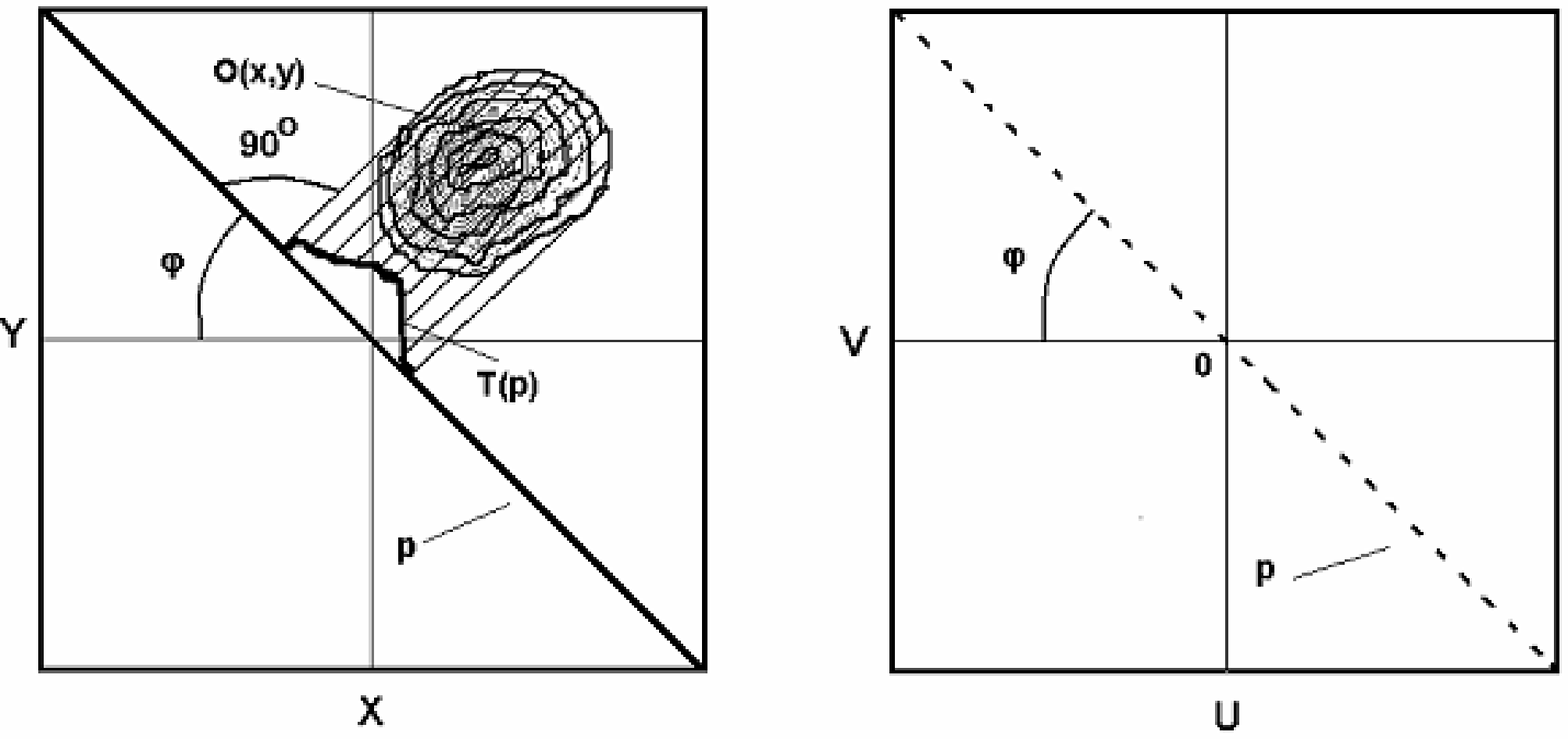}
 \end{center}}
\begin{center}\hskip 0cm  (a) \hskip 4.5 cm (b) \end{center}
\begin{center} {\bf Fig.~1.} Graphical illustration of the theory of projections: (a) region in which the object is defined $(X,Y)$; (b) region of spatial
frequencies of the object $(U,V)$.
\end{center}
\end{figure}

\newpage
\begin{figure}[t]
{\begin{center}
  \includegraphics[width= 100mm]{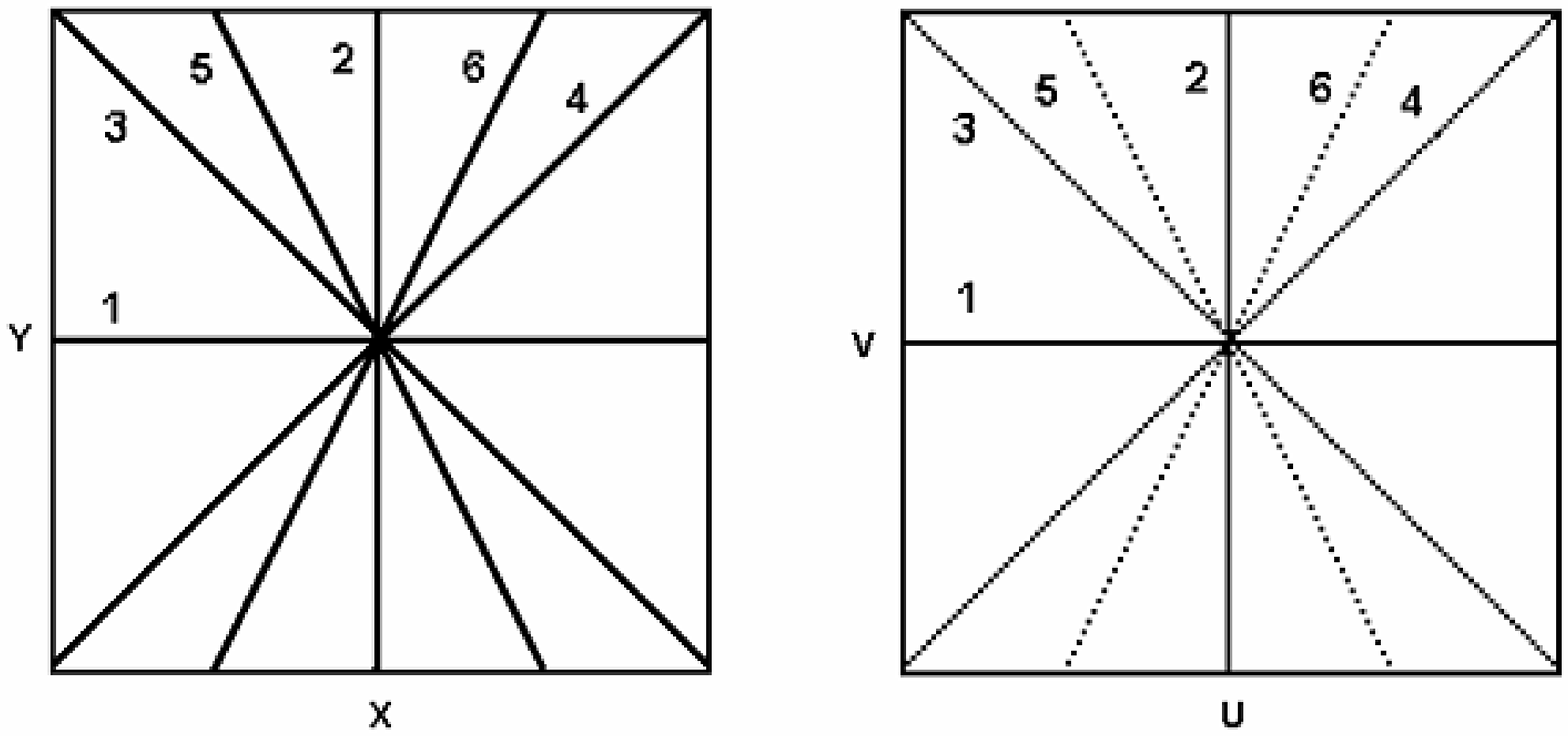}
\end{center}}
\begin{center}\hskip 0cm  (a) \hskip 4.5 cm (b) \end{center}
\begin{center}{\bf Fig.~2.} Example of six projections: (a) geometry of the projections in the region of the object $(X,Y)$; (b) corresponding coverage
of the spatial-frequency domain for the object $(U,V)$. The
numbers denote the numbers of the projections.
\end{center}
 \end{figure}

\newpage
\begin{figure}[t]
{\begin{center}
   \includegraphics[width= 100mm]{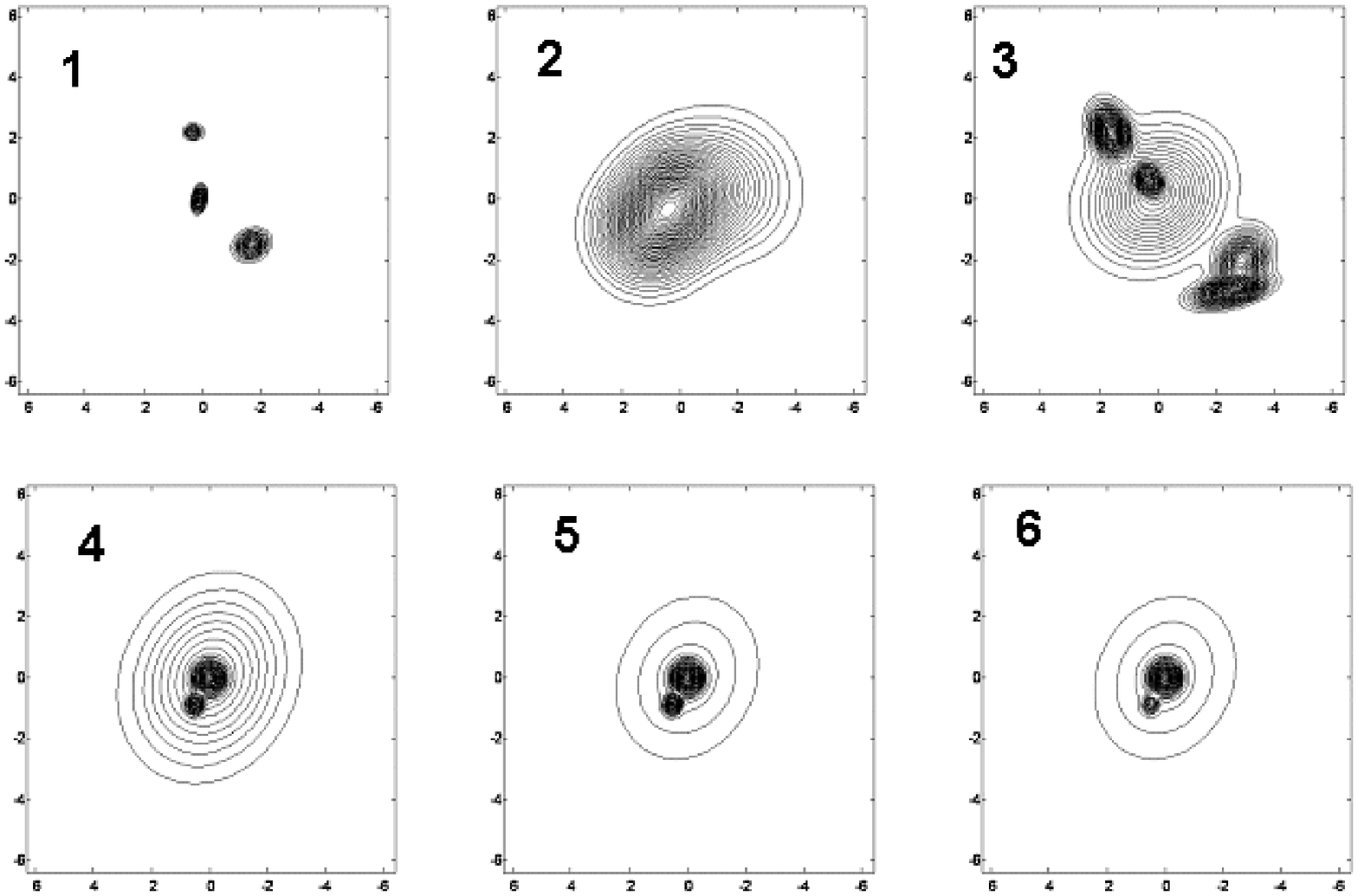}
\end{center}}
\begin{center}
{\bf Fig.~3.} Images of model radio sources with various
morphological types. The numbers {\it 1–6} denote the source
number.
\end{center}
 \end{figure}

\newpage
\begin{figure}[t]
{\begin{center}
   \includegraphics[width= 100mm]{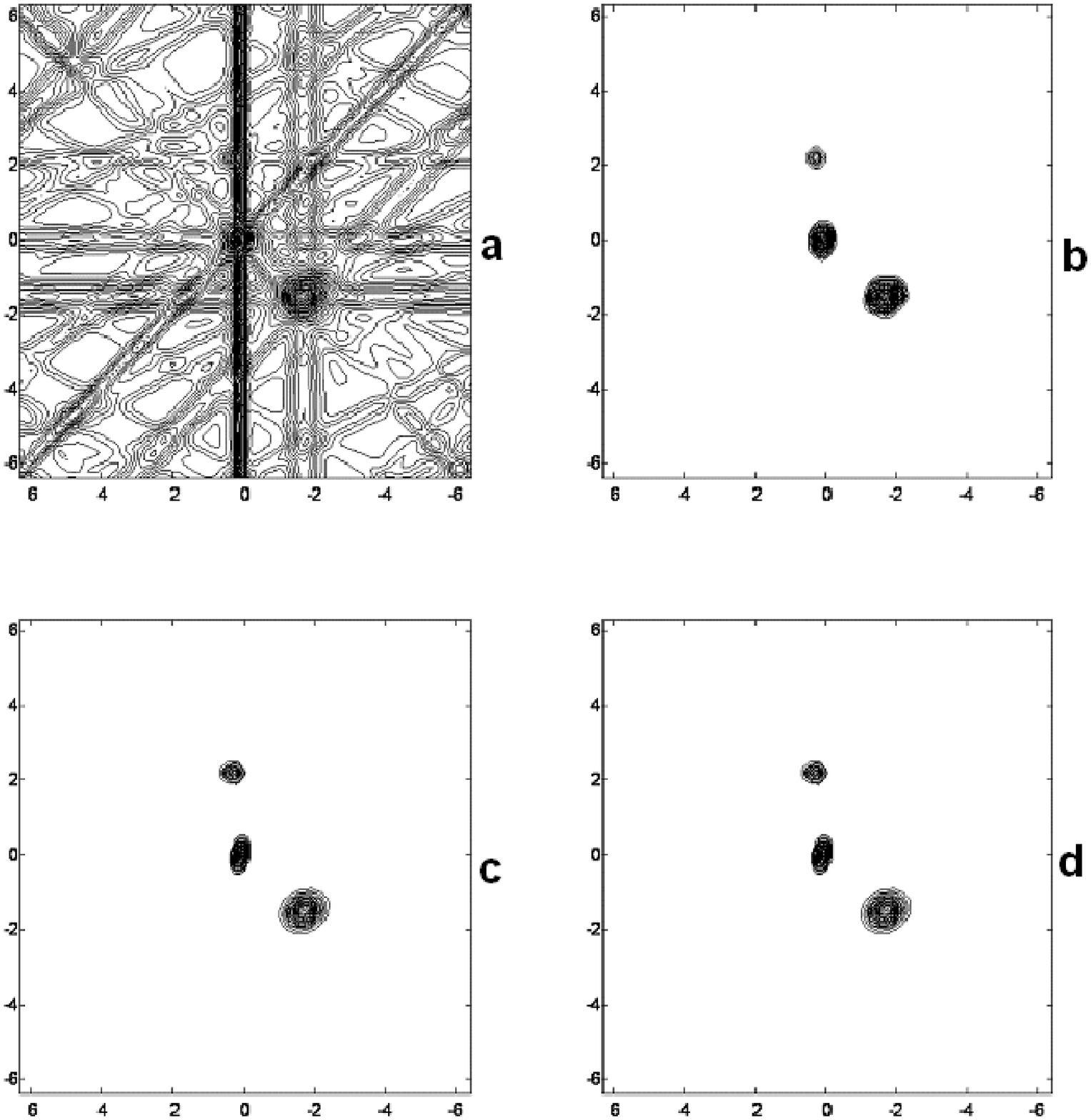}
\end{center}}
\begin{center}
{\bf Fig.~4.} Results of reconstructing radio source {\it 1}: (a)
the dirty image, (b) the CLEAN image, (c) the MEM image and (d)
the image obtained using the difference generalized MEM algorithm.
\end{center}
 \end{figure}

\newpage
\begin{figure}[t]
{\begin{center}
   \includegraphics[width= 100mm]{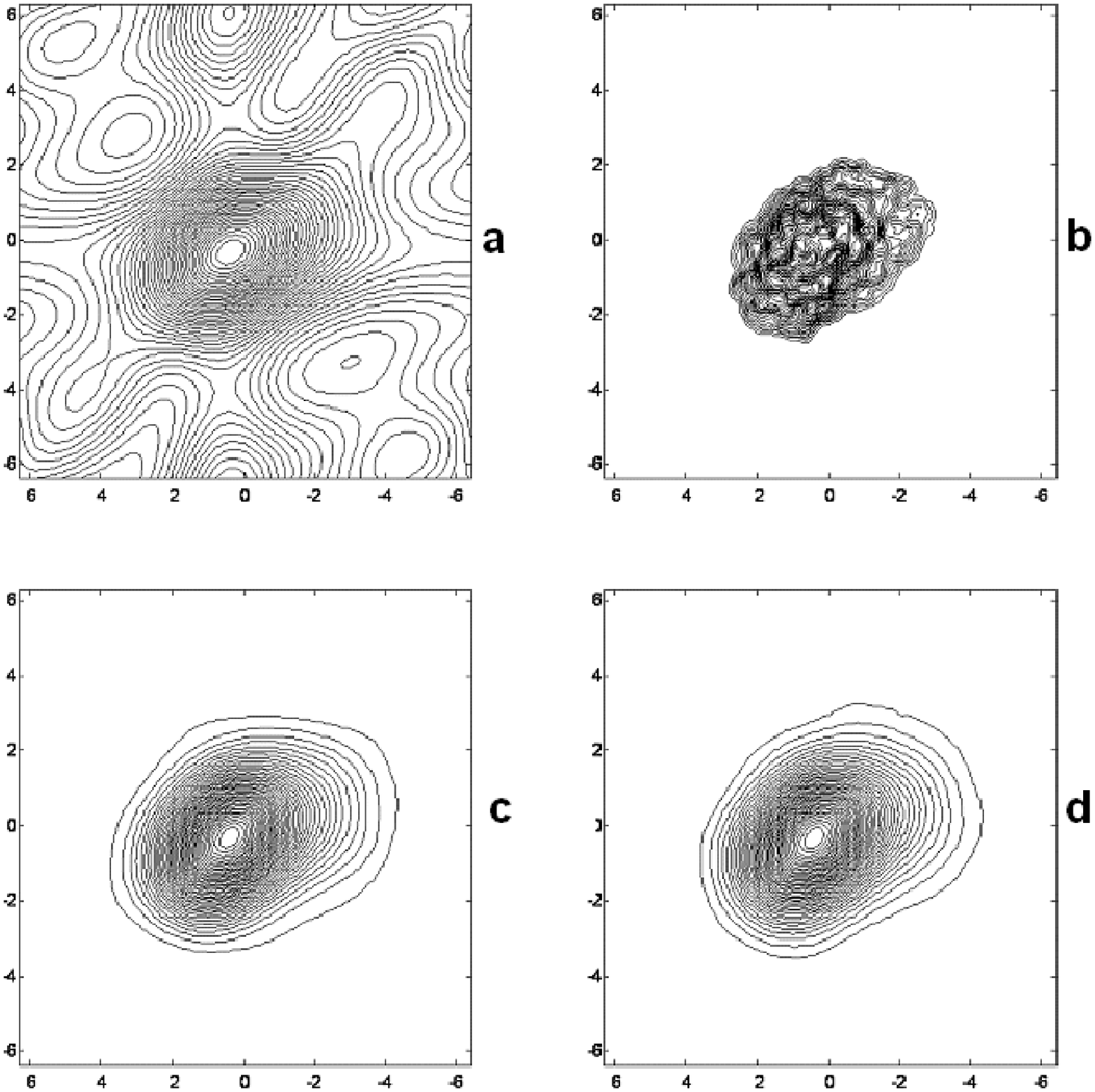}
\end{center}}
\begin{center}
{\bf Fig.~5.} Same as Fig. 4 for radio source {\it 2}.
\end{center}
 \end{figure}

\newpage
\begin{figure}[t]
{\begin{center}
             \includegraphics[width= 100mm]{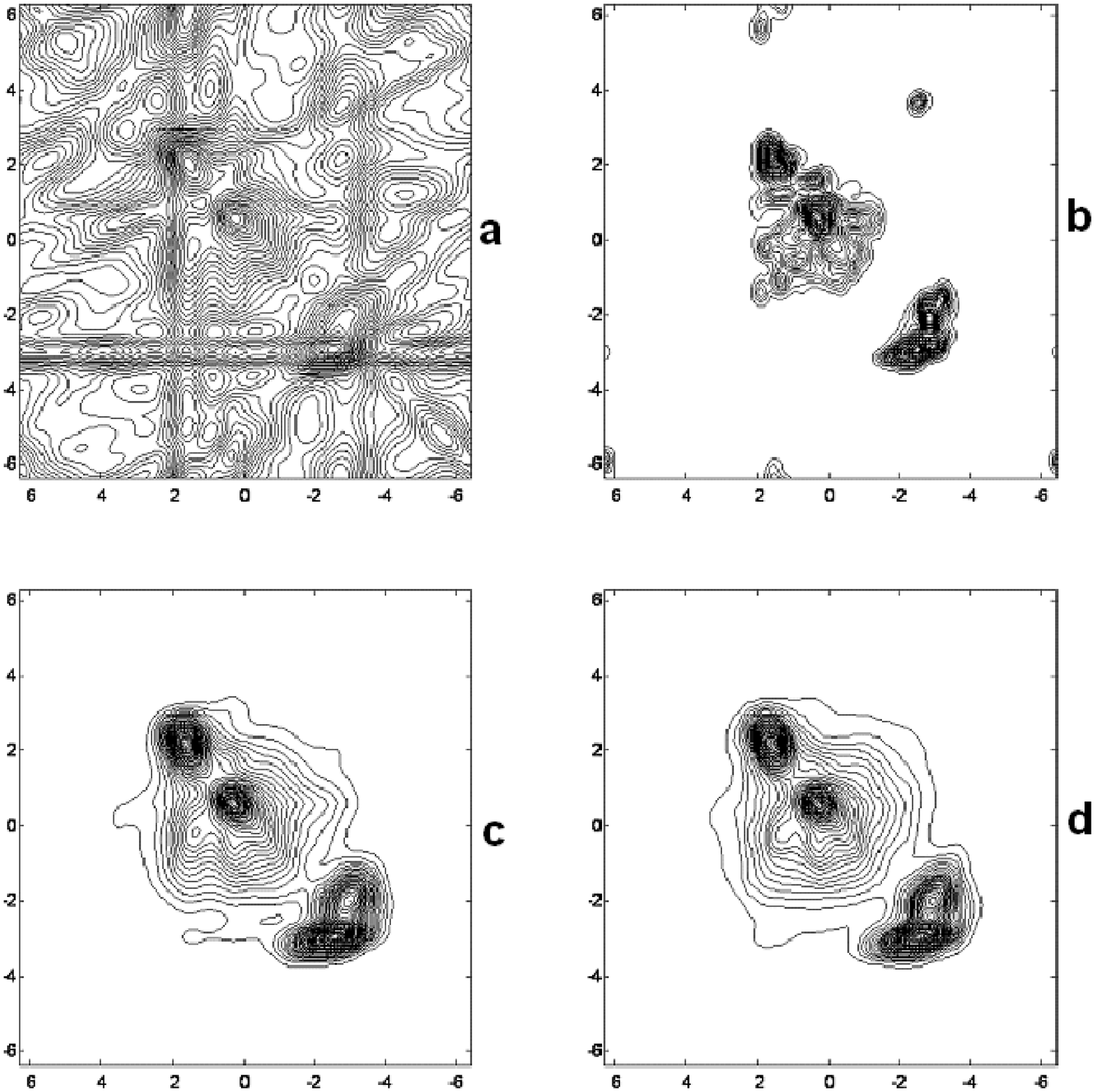}
\end{center}}
\begin{center}
{\bf Fig.~6.} Same as Fig. 4 for radio source {\it 3}.
\end{center}
 \end{figure}

 \newpage
\begin{figure}[t]
{\begin{center}
             \includegraphics[width= 100mm]{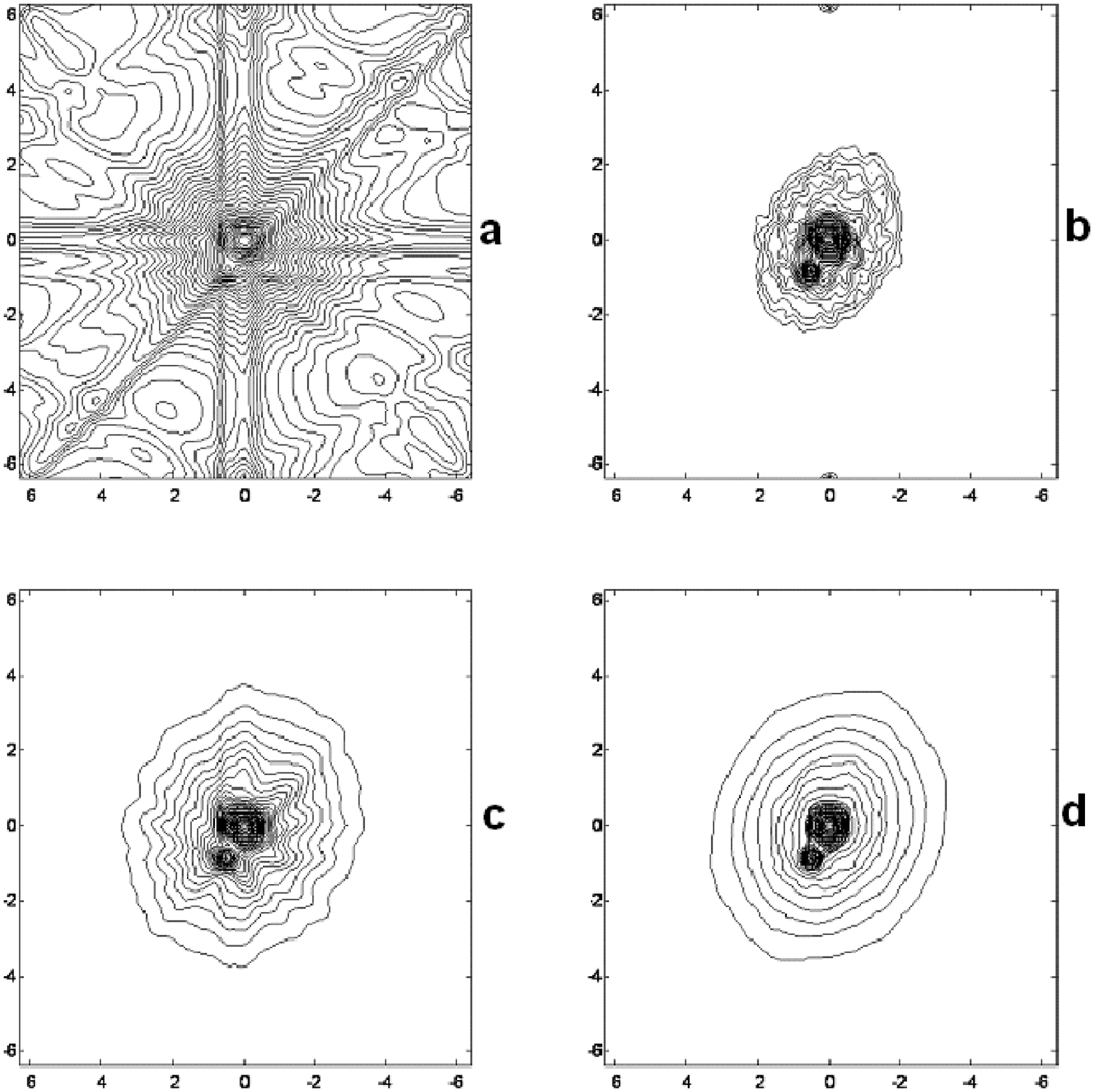}
\end{center}}
\begin{center}
{\bf Fig.~7.} Same as Fig. 4 for radio source {\it 4}.
\end{center}
 \end{figure}

 \newpage
\begin{figure}[t]
{\begin{center}
             \includegraphics[width= 100mm]{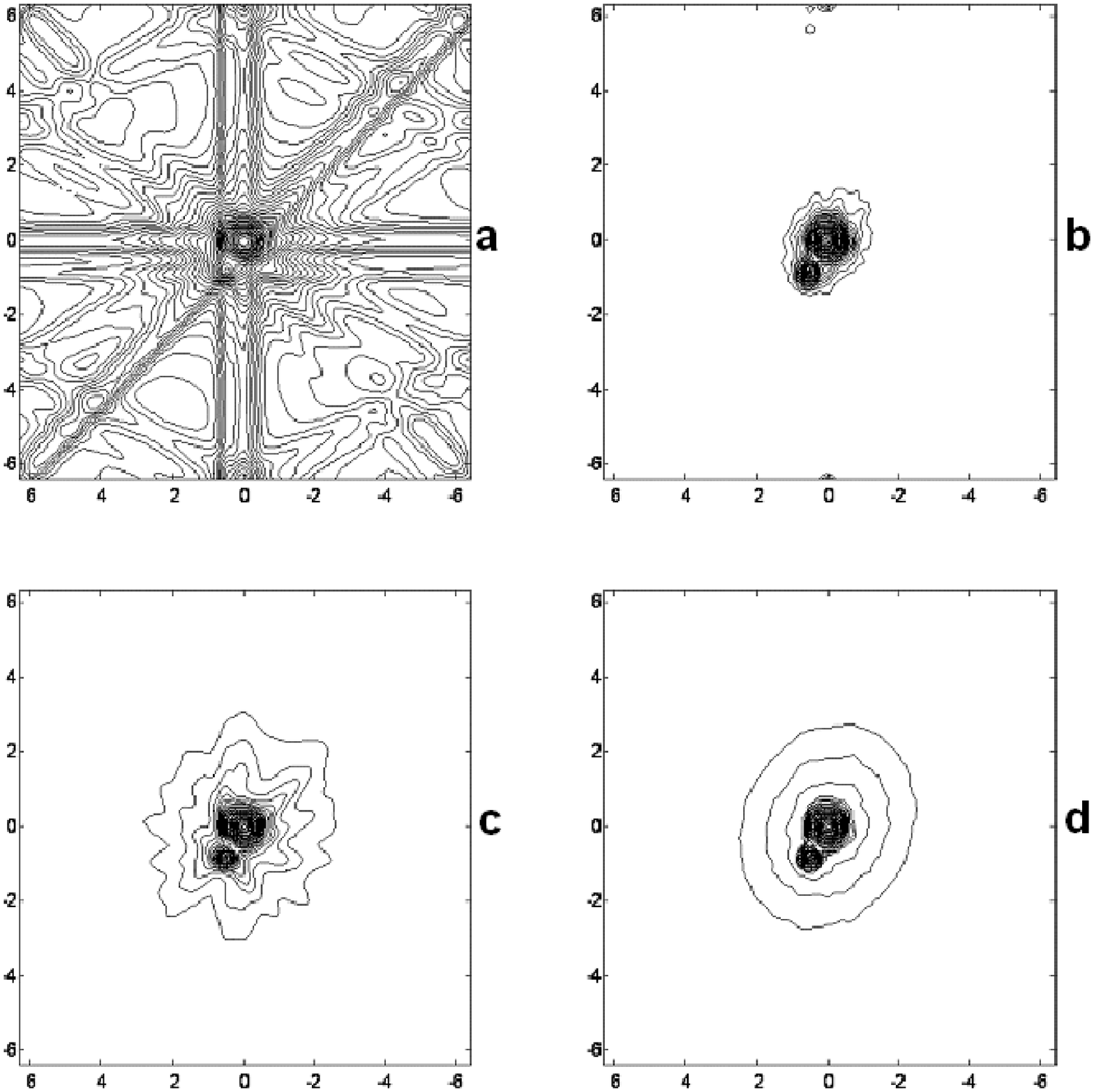}
\end{center}}
\begin{center}
{\bf Fig.~8.} Same as Fig. 4 for radio source {\it 5}.
\end{center}
 \end{figure}

 \newpage
\begin{figure}[t]
{\begin{center}
             \includegraphics[width= 100mm]{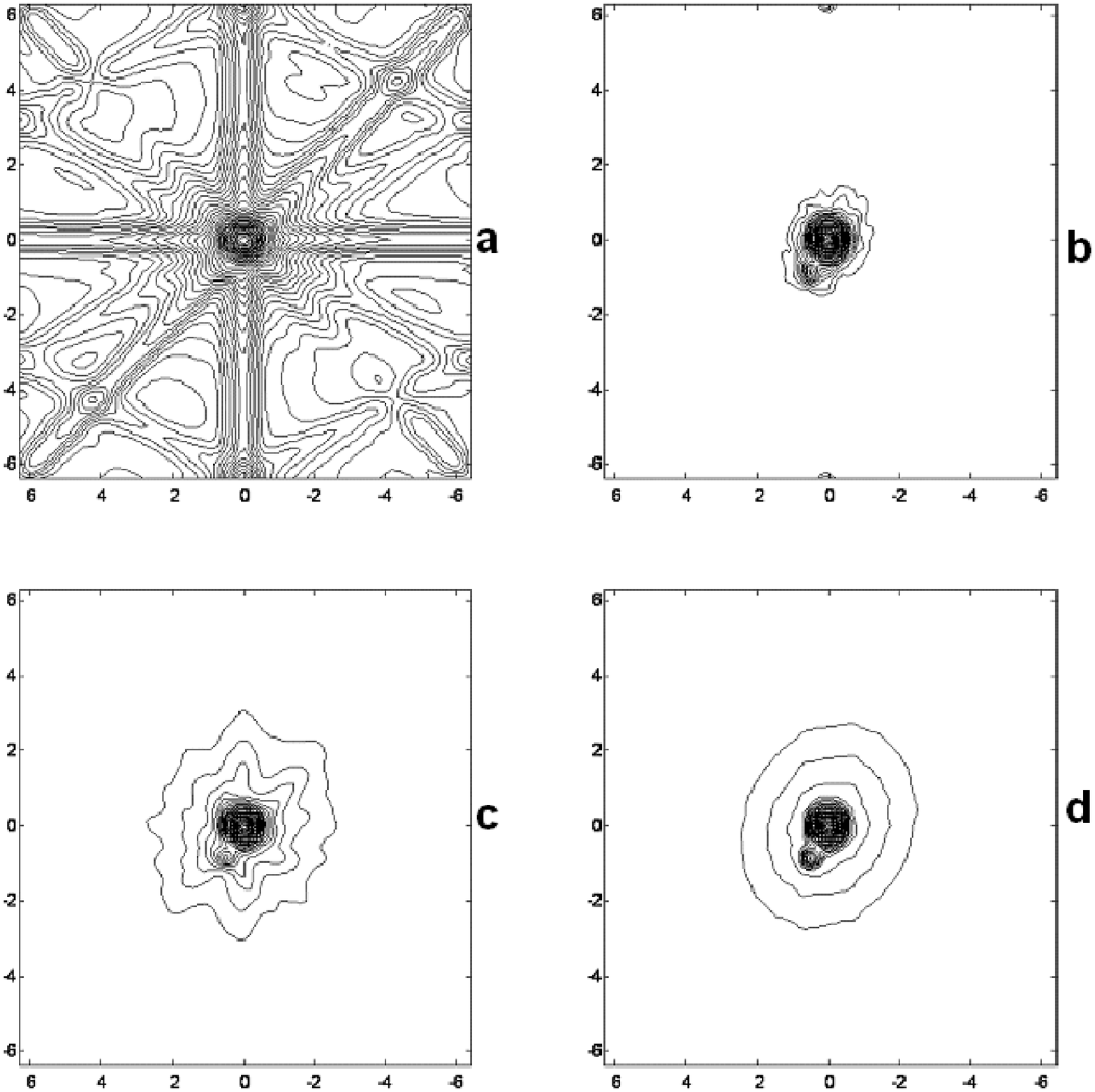}
\end{center}}
\begin{center}
{\bf Fig.~9.} Same as Fig. 4 for radio source {\it 6}.
\end{center}
 \end{figure}

\end{document}